\documentclass[fleqn,usenatbib, useAMS]{mnras}

\usepackage{sansmath}
\usepackage{makecell}
\usepackage{color}
\usepackage{xcolor}

\usepackage{bm,cancel}
\usepackage{booktabs}
\usepackage[T1]{fontenc}

\DeclareRobustCommand{\VAN}[3]{#2}
\let\VANthebibliography\thebibliography
\def\thebibliography{\DeclareRobustCommand{\VAN}[3]{##3}\VANthebibliography}

\usepackage{graphicx}	% Including figure files
\usepackage{amsmath}	% Advanced maths commands
\usepackage{amssymb}	% Extra maths symbols

\usepackage{soul}
\usepackage{hyperref}
\def\gp{\gamma_+}
\def\gc{\gamma_\times}
\def\eint{e_{\mathrm{int}}}
\def\eobs{e_{\mathrm{obs}}}
\def\thetaobs{\theta_{\mathrm{obs}}}
\usepackage{scalerel}
\usepackage{tikz,xcolor}
\usetikzlibrary{svg.path}
\definecolor{lime}{HTML}{A6CE39}
\DeclareRobustCommand{\orcidicon}{%
	\begin{tikzpicture}
	\draw[lime, fill=lime] (0,0) 
	circle [radius=0.16] 
	node[white] {{\fontfamily{qag}\selectfont \tiny ID}};
	\draw[white, fill=white] (-0.0625,0.095) 
	circle [radius=0.007];
	\end{tikzpicture}
	\hspace{-2mm}
}
\foreach \x in {A, ..., Z}{%
	\expandafter\xdef\csname orcid\x\endcsname{\noexpand\href{https://orcid.org/\csname orcidauthor\x\endcsname}{\noexpand\orcidicon}}
}

\title{Kinematic Lensing Inference \MakeUppercase{\romannumeral 1}: Characterizing Shape Noise with Simulated Analyses}

\author[Pranjal R. S. et al.]{Pranjal R. S.\orcidA{},$^{1}$\thanks{E-mail: pranjalrs@arizona.edu}
Elisabeth Krause\orcidB{},$^{1, 2}$
Hung-Jin Huang,$^1$
Eric Huff,$^3$
Jiachuan Xu\orcidC{},$^1$
\newauthor
Tim Eifler,$^1$ 
Spencer Everett${^3}$
\\
$^{1}$Department of Astronomy/Steward Observatory, University of Arizona, 933 North Cherry Avenue, Tucson, AZ 85721, USA \\
$^{2}$Department of Physics, University of Arizona, 1118 E Fourth Str, AZ 85721, USA\\
$^{3}$Jet Propulsion Laboratory, California Institute of Technology, Pasadena, CA 91109, USA
}

\date{Accepted XXX. Received YYY; in original form ZZZ}

\pubyear{2015}

\begin{document}
\label{firstpage}
\pagerange{\pageref{firstpage}--\pageref{lastpage}}
\maketitle

\begin{abstract}
The unknown intrinsic shape of source galaxies is one of the largest uncertainties of weak gravitational lensing (WL). It results in the so-called shape noise at the level of $\sigma_\epsilon^{\mathrm{WL}} \approx 0.26$, whereas the shear effect of interest is of order percent.
Kinematic lensing (KL) is a new technique that combines photometric shape measurements with resolved spectroscopic observations to infer the intrinsic galaxy shape and directly estimate the gravitational shear. This paper presents a KL inference pipeline that jointly forward-models galaxy imaging and slit spectroscopy to extract the shear signal. We build a set of realistic mock observations and show that the KL inference pipeline can robustly recover the input shear. To quantify the shear measurement uncertainty for KL, we average the shape noise over a population of randomly oriented disc galaxies and estimate it to be $\sigma_\epsilon^{\mathrm{KL}}\approx 0.022-0.038$ depending on emission line signal-to-noise. This order of magnitude improvement over traditional WL makes a KL observational program feasible with existing spectroscopic instruments. To this end, we characterize the dependence of KL shape noise on observational factors and discuss implications for the survey strategy of future KL observations. In particular, we find that prioritizing quality spectra of low inclination galaxies is more advantageous than maximizing the overall number density.

%In future work, we will build on the pipeline presented here, accounting for complex galaxy morphology and kinematic structure,  with the goal of measuring the KL signal from real data.
\end{abstract}


%%%%%%%%%%%%%%%%%%%%%%%%%%%%%%%%%%%%%%%%%%%%%%%%%%

%%%%%%%%%%%%%%%%% BODY OF PAPER %%%%%%%%%%%%%%%%%%

\section{Introduction}

Following the first detections of cosmic shear \citep{Wittman_2000} and galaxy-galaxy lensing \citep{Brainnerd_1996, Dell_Antonio_1996} more than two decades ago, weak lensing (WL) has matured into an important cosmological probe with recent measurements using millions of galaxies to place tight constraints on cosmological parameters \citep{van_Uitert_2018, Hikage_2019, Hamana_2020, KiDS_cosmic_shear, DESY3_Secco2022, DESY3_Amon2022}. High precision measurements of galaxy-galaxy lensing have also enabled studies of the galaxy-halo connection \citep{KiDS_ggl, HSC_ggl, KiDS_ggl2, DESY3_Zacharegkas2021}, which is not only important in a cosmological context but also for understanding galaxy evolution.

The coming decade is set to be even more data-rich and exciting for WL with several upcoming Stage-IV programs. Wide-field surveys like the Rubin Observatory’s Legacy Survey of Space and Time (LSST\footnote{\href{https://www.lsst.org}{\nolinkurl{https://www.lsst.org}}}, \citealt{LSST}), \textit{Nancy Grace Roman Space Telescope} (\textit{Roman}\footnote{\href{https://roman.gsfc.nasa.gov}{\nolinkurl{https://roman.gsfc.nasa.gov}}}, \citealt{Roman}) and \textit{Euclid}\footnote{\href{https://sci.esa.int/web/euclid}
{\nolinkurl{https://sci.esa.int/web/euclid}}} \citep{Euclid} will significantly enhance the measurement precision and further improve cosmological constraints.

Extensive efforts go into modeling, calibrating, and testing WL systematics. For example, on the observational side, accurate modeling of the point-spread function (PSF) \citep{Jarvis_2020} and effects like blending, selection effects, and detector non-idealities have to be taken into consideration \citep{MasseySystematics, Cropper_2013, Bernstein_2017,MandelbaumLensingReview, Choi_2020, Hirata_2020} before the response of the estimator to the underlying shear is calibrated \citep{HScalibration,Huff_2017, Sheldon_2017, DESY1_Zuntz, Sheldon_2020}. Redshift uncertainties can bias cosmological inference if the redshift distribution of the source sample is not accurately determined \citep{Bernstein_2010, KiDS+V_Hildebrandt, DESY3_Myles21}. Astrophysical uncertainties such as intrinsic alignments \citep{Catelan_IA, Hirata_IA, Joachimi_2015, Krause_2016, Blazek_19, DESY3_Secco2022} and baryonic effects that can redistribute matter on small scales \citep{van_Daalen_2011, Zentner_2013, Mead_2015, Eifler_2015, Chisari_2019, Schneider_2019, Huang_2019, Huang_2021} also have to be accounted for.

Significant progress has been made in modeling and mitigating these systematic effects; however, WL measurements remain a statistical challenge as the shear signal is an order of magnitude smaller than the dispersion in the intrinsic galaxy shapes, the so-called shape noise $\sigma_\epsilon$. Consequently, lensing measurements have to include low signal-to-noise galaxies to increase the statistical precision, which is the main cause of the systematics complexities.

Alternatively, one can reduce the level of shape noise by inferring the intrinsic galaxy shape separately from the shear effect. To illustrate the degeneracy between intrinsic shape and shear, we express ellipticities as complex numbers, 
$\mathbf{\epsilon} = \epsilon_+ + i\epsilon_\times$, with modulus equal to the scalar ellipticity $e=|\mathbf{\epsilon}|$. Here the components $+$ and $\times$ are aligned with the major axis of the galaxy and rotated by $45^\circ$ with respect to the major axis, respectively. For a weak shear $\gamma = \gp+i\gc$, the relationship between the intrinsic ellipticity $\eint$ and observed ellipticity $\eobs$ is
\begin{align}
        \epsilon_{\mathrm{obs}} = \epsilon_{\mathrm{int}} +\gamma.
        \label{eq:eobs}
\end{align}
 Traditional shear measurement methods use observed galaxy shapes as a proxy for shear and take their ensemble average, assuming galaxies to be randomly oriented, i.e., $\langle \epsilon_{\mathrm{int}}\rangle=0$. The variance of such shear estimators is dominated by shape noise, $\mathrm{Var}(\hat{\gamma}_{+,\times})= \sigma_\epsilon^2/N$ with $\sigma_\epsilon=\left(\langle \epsilon_{\mathrm{int},+}^2+ \epsilon_{\mathrm{int},\times}^2\rangle/2\right)^{1/2}\approx 0.26$ being the component-wise ellipticity dispersion and $N$ the number of galaxies. Hence precise measurements of percent-level shear require large galaxy samples.

Several methods have been proposed to break the degeneracy between shape and shear. For example, \cite{Blain_2002} and \cite{Morales_2006} proposed the use of projected galaxy velocity fields to infer intrinsic galaxy shapes:
In the absence of lensing, the axes along which maximum and zero circular velocity occur, i.e., the kinematic major and minor axes, are orthogonal. As shown in Fig. \ref{fig:shear_illustration}, this is no longer the case in the presence of shear, which breaks the axisymmetry in the velocity field. 
Thus with sufficiently resolved 2D spectroscopic observations, it is possible to measure shear from disc kinematics. \cite{deBurgh-Day_2015} built on these ideas and implemented a technique to recover shear by looking for asymmetries in IFU data, thus utilizing both shape and velocity information. \cite{Gurri2020} obtained the first measurements of the shear signal using velocity maps from a sample of 18 low-\textit{z} galaxy-galaxy lensing systems, reporting an average shear of $\langle\gamma\rangle=0.020\pm0.008$. Adding shape information from galaxy imaging to this method can improve the statistical error by factors of $2\textrm{--}6$ on the inferred shear as shown by \cite{DiGiorgio_2021}.

This work builds on the kinematic lensing (hereafter KL) methodology presented in \citealt{huff2019} (also see \citealt{Xu_2022}). KL utilizes spatially resolved spectroscopy of disc galaxies along with the Tully-Fisher relation (TFR, \cite{TFR_1977}) to reduce shape noise. In addition, KL is also robust to three key systematic uncertainties of traditional weak lensing measurements: 1) Redshift uncertainties are eliminated with resolved spectra for each source galaxy. 2) The galaxy shape inferred from kinematics is the unlensed galaxy shape, which includes intrinsic alignments (IA). 3) The reduction in shape noise allows a KL measurement to focus on bright, well-resolved, relatively isolated galaxies where photometric shape measurement biases are typically very small.

This work represents the first in a series of publications building up to measuring a KL signal from data. We build realistic mock observations and develop a KL inference pipeline that estimates shear from a joint analysis of galaxy imaging and slit spectroscopy. We test and validate our pipeline for the specific case of high-resolution imaging and Keck DEIMOS slit spectra and estimate a realistic level of KL shape noise for such a measurement.

The paper is organized as follows: In section~\ref{sec:kl_theory}, we review the effects of lensing on galaxy photometry and kinematics and show how the latter can be used to estimate shear. In section~\ref{sec:KL_obs}, we describe the simulation code for producing mock observations and the fast forward model used for parameter inference. We present our inference procedure and results in section~\ref{sec:kl_measurement}. We explore the dependence of KL shape noise on galaxy properties and survey strategy in section \ref{sec:discussion} and conclude in section \ref{sec:conclusion}.

\section{Kinematic Lensing}
\label{sec:kl_theory}
\begin{figure}
  \centering
  \includegraphics[width=\linewidth]{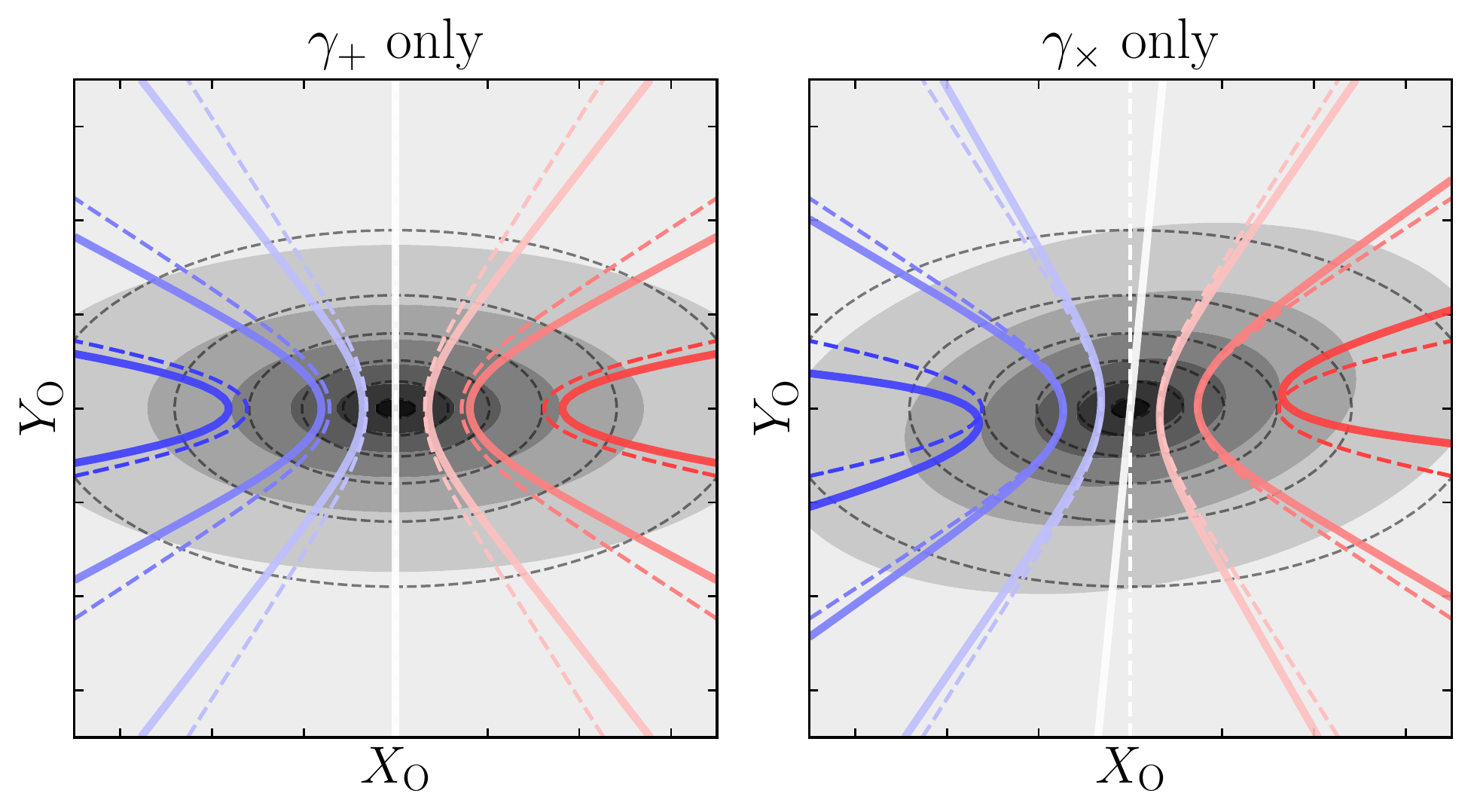}
  \caption{Illustration of the effects of shear on galaxy photometry and the projected velocity field. The left and right panels correspond to the application of the $\gp$ and $\gc$ components. In both panels, the original and observed shapes are shown as unfilled dashed elliptical contours and filled elliptical contours, respectively. Similarly, the original and observed velocity fields are shown as dashed blue-red contours and solid blue-red contours, respectively.}
  \label{fig:shear_illustration}
\end{figure}

Weak gravitational distorts the shapes of background galaxies, which can be understood as a transformation that maps a galaxy image from the source plane $X_S\textrm{--}Y_S$ to the image plane $X_I\textrm{--}Y_I$. In the linear regime, this mapping is expressed in terms of the lensing distortion matrix \textsf{\textbf{A}}
\begin{equation}\label{eq:lensing_mat}
    \textsf{\textbf{A}} = 
\begin{pmatrix} 
1-\gp & -\gc \\
-\gc & 1 + \gp
\end{pmatrix}.
\end{equation}
\textsf{\textbf{A}} is defined to be the transformation that reverses the effect of WL and maps a position vector in the image plane $\boldsymbol{x}_I = (x_I, y_I)^\mathrm{T}$ to the position vector in source plane $\boldsymbol{x}_S = (x_S, y_S)^\mathrm{T}$ i.e, $\boldsymbol{x}_S = \textsf{\textbf{A}} \cdot \boldsymbol{x}_I$. 

The shear components $\gp$ and $\gc$ in Eq. (\ref{eq:lensing_mat}) describe the stretching of the galaxy caused by lensing as shown in Fig. \ref{fig:shear_illustration}. In summary, for a galaxy with intrinsic ellipticity $\eint$ lensing distorts the observed ellipticity $\eobs$ and position angle $\thetaobs$ as \citep{BernsteinJarvis_02,huff2019}
\begin{align}
\label{eq:e_obs}
\eobs &= \eint +2\gp(1-\eint^2),\\
\label{eq:theta_obs}
    \thetaobs &= \frac{\gc}{\eint},
\end{align}
where the second equation assumes an intrinsic position angle $\theta_{\mathrm{int}}=0$.

In order to break this degeneracy, we incorporate the characteristic signature of lensing on the line-of-sight (LoS) velocity field of disc galaxies, as illustrated in Fig. \ref{fig:shear_illustration}.
The $\gc$ component misaligns the kinematic and photometric axes. As a result, the photometric minor axis no longer corresponds to a zero LoS velocity\footnote{For simplicity, we assume the axes of the unlensed galaxy to be aligned with the source plane. For a derivation in a more general coordinate system, we refer to Appendix A in \citealt{Xu_2022}.}. The LoS velocity along the photometric major axis is largely unaffected by lensing (up to first order in shear). Analytically, these relations for the photometric minor axis velocity $ v'_{\mathrm{minor}}$ and major axis velocity $v'_{\mathrm{major}}$ can be expressed as \citep{huff2019}
\begin{align}
    v'_{\mathrm{minor}} &= \gc v_{\mathrm{circ}} \cos i \sqrt{\frac{2(1+\eint)}{(1-q_z^2)\eint}}, \label{eq:vminor}\\
    v'_{\mathrm{major}} &= v_{\mathrm{circ}} \sin i, \label{eq:vmajor}
\end{align}
where $i$ is the inclination angle, $q_z$ is the edge-on aspect ratio of the galaxy, and $v_{\mathrm{circ}}$ is the maximum circular velocity.

The estimate of the maximum circular velocity $\hat{v}_{\mathrm{circ}}$ is obtained from the TFR which relates it to the galaxy's absolute magnitude $M_\mathrm{B}$
\begin{align}
\label{eq:TFR}
    \log \hat{v}_{\mathrm{circ}} = a(M_\mathrm{B} + M_\mathrm{p}) + b,
\end{align}
where $M_\mathrm{p}$ is the pivot value, $a$ is the slope and $b$ is the intercept. The scatter in the measured TFR depends on the sample redshift, photometric band, etc, and is usually in the range $\sigma_\mathrm{TF}=0.05-0.12$ dex \citep{Chiu_2007, Miller_2011, Reyes_2011}. 

Combining $\hat{v}_{\mathrm{circ}}$ predicted  by the TFR and the measured photometric major axis velocity $v'_{\mathrm{major}}$ allows us to infer the underlying disc inclination $\sin i$ by using Eq. (\ref{eq:vmajor}). The disc inclination can then be related to the intrinsic ellipticity $\eint$ using a geometric relation
\begin{align}
    \eint = \frac{(1-q_z^2) \sin^2 i}{2 - (1-q_z^2)\sin^2 i}.
\end{align}
With an estimate of the intrinsic galaxy shape, one can then infer $\gp$ using Eq. (\ref{eq:e_obs}), where the observed shape is measured from the photometry of the disc.
Thus, for disc galaxies combining photometry with galaxy kinematics and the TFR allows us to distinguish shear from intrinsic galaxy properties (shape and position angle).

While the simple estimators derived here are helpful to illustrate how KL combines different types of measurement, in practice, forward modeling of the observables (velocity field and photometry) will be better suited for KL shear measurements. Effects like point-spread function (PSF) smearing, non-idealized galaxy morphology, and observational effects like finite slit size or slit offsets and misalignment are relatively easy to incorporate in a forward model of the observables but nearly intractable in the estimator approach.

\section{Simulating KL observations}
\begin{figure*}
  \centering
  \includegraphics[width=\linewidth]{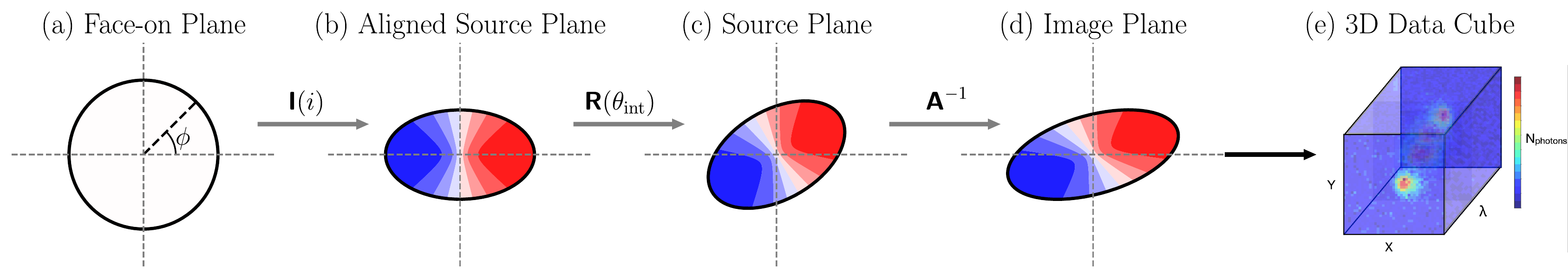}
  \caption{Coordinate transformations performed to compute the observed spectrum data cube. The galaxy isophote is shown as a solid black line and the LoS velocity as blue-red contours. Panel (a): In the face-on plane, the galaxy is a circular disc with a zero LoS velocity component. Panel (b): The circular disc is inclined to account for the galaxy's intrinsic ellipticity, which also introduces a LoS velocity field. Panel (c): The inclined disc is rotated to account for the galaxy's intrinsic position angle. Panel (d): The LoS velocity field in the image plane is obtained by applying the shear transformation. Panel (e): Combining the LoS velocity field with the emission line profile and galaxy photometry results in a 3D data cube which is then used to compute the 2D spectrum.}
\label{fig:tfcube_overview}
\end{figure*}

\label{sec:KL_obs}
The first step in estimating shear using KL is to model the effects of lensing on galaxy observables, i.e., photometry and kinematics. 
In this work, we infer the kinematic structure of the galaxy from long-slit spectroscopy, although a similar methodology can be developed for grism or IFU observations.

We use two separate simulation codes, one for generating mock data and the other for parameter inference. Referred to as the mock generator and the fast forward model, respectively, the basic physical recipe for modeling the spectrum in both these schemes is the same. The key difference is that the mock data generator is designed to simulate properties of real data, taking into account several observational effects, whereas the fast forward model is optimized for computational speed.

\subsection{Mock Image Generator}
\label{sec:image_model}

We use the open-source software \texttt{Galsim} \citep{RJM+15} for modeling galaxy images. \texttt{Galsim} can generate images from a variety of parametric models and provides routines for applying effects such as shear and rotation, as well as several PSF and noise models. \texttt{Galsim} was developed for simulating weak lensing shears, and its extensive use and testing in several studies ensures that operations like shear transformations and convolutions are performed with an accuracy sufficient for our purpose.

For the photometry of the disc, we assume a $n=1$ S\'ersic profile with half-light radius $r_{\mathrm{hl}}^{\mathrm{image}}$ and use the \textsc{InclinedSersic} object to create a disc with inclination $\sin i$ and edge-on aspect ratio $q_z$. The disc is then rotated to account for the intrinsic galaxy orientation, followed by the application of two shear components and a convolution with the seeing, which is a combination of PSFs from the atmosphere and the instrument being modeled. Finally, we add Gaussian noise for the specified image signal-to-noise.

Due to the low computational cost and ease of implementation, we also use \texttt{Galsim} for generating images in the fast forward model.

\subsection{Mock Spectrum Generator}
\label{section:tfCube}
In the coordinate frame of the source galaxy, we model it as an intrinsically round disc with a circularly symmetric rotation velocity profile given by the $\tan^{-1}$ function \citep{Courteau_1994, Green_2014}. We project the azimuthal velocity directions into the LoS directions, depending on the inclination ($\sin i$) and the polar angle ($\phi$, as measured from the galaxy's major axis in the face-on frame)
\begin{equation}
\label{eq:vel_profile}
    v(r) = v_0 + \frac{2}{\pi} v_{\mathrm{circ}} \cos (\phi) \sin (i) \tan^{-1}\left(\frac{r}{r_{\mathrm{vscale}}}\right),
\end{equation}
where $v_0$ is the galaxy systemic velocity, $v_{\mathrm{circ}}$ is the maximum circular velocity, $r_{\mathrm{vscale}}$ is the velocity scale radius.

We apply a shear distortion (i.e., the inverse of the lensing distortion matrix $\textsf{\textbf{A}}$ in Eq. (\ref{eq:lensing_mat})) on the two-dimensional source plane to transform the velocity grid of the source galaxy to the image frame. These transformations, illustrated in panels (a) through (d) in Fig. \ref{fig:tfcube_overview}, can be summarized as

\begin{equation}
\label{eq:coord_mapping}
\boldsymbol{x}_I= \textsf{\textbf{A}}^{-1}\text{ }\textsf{\textbf{R}}\text{ }\textsf{\textbf{I}}\text{ }\boldsymbol{x}_{\mathrm{face-on}},
\end{equation}

where $\boldsymbol{x}_{\mathrm{face-on}}$ is a position vector in the face-on plane, $\textsf{\textbf{R}}$ is a rotation matrix and $\textsf{\textbf{I}}$ accounts for galaxy inclination. The latter two are given by
\begin{equation}
\textsf{\textbf{R}}(\theta) = 
\begin{pmatrix} 
\cos \theta & -\sin \theta \\
\sin \theta & \cos \theta
\end{pmatrix}; \hspace{5pt}
\textsf{\textbf{I}}(i) = 
\begin{pmatrix} 
1 & 0 \\
0 & \cos i
\end{pmatrix}.
\end{equation}
Next, we combine the distorted 2D photometry (as described in Section \ref{sec:image_model}) together with the LoS velocity information to form a 3D data cube, as shown in the right panel of Fig.~\ref{fig:tfcube_overview}. 
For each spatial position ($x_I$, $y_I$), we
assign a Gaussian emission line profile with the peak at the wavelength
\begin{equation}
\lambda(x_I, y_I) = (1+z)(1 + \frac{v(x_I, y_I)+v_0}{c})\lambda_0, 
\end{equation}
where $\lambda_0$ is the rest-frame emission line wavelength, $z$ is the galaxy redshift, $v_0$ is the systemic velocity, and $c$ is the speed of light. The amplitude of the Gaussian is assigned following the 2D spatial photometry and then normalized so that the average spectral line intensity within a given mask (e.g., fiber or slit mask) is matched with a reference SDSS galaxy at a similar redshift.
The variance of the Gaussian is
\begin{equation}
\label{eq:sigma2}
    \sigma^2 = \frac{(1+z) \lambda_0}{R^2} + \sigma_{\rm int}^2 \, ,
\end{equation} 
where $R$ is the spectral resolution of the spectrograph ($R=\lambda/\Delta\lambda$), and $\sigma_{\rm int}^2$ is the intrinsic velocity dispersion along the LoS for the disc galaxy. For simplicity, we set $\sigma_{\rm int}^2=0.01$ nm at all spatial positions as the total velocity dispersion is dominated by the instrumental contribution, the first term Eq. \eqref{eq:sigma2}.

To generate mock data with various observational effects, we take into account sky emissions, atmospheric transmission, and the effect of PSF. The input sky template is calculated using the \texttt{SkyCalc}\footnote{\url{https://www.eso.org/observing/etc/bin/gen/form?INS.MODE=swspectr+INS.NAME=SKYCALC}} tool \citep{Noll_2012, Jones_2013} and
the integrated intensity is multiplied by the atmosphere transmission at each wavelength. To inject noise, we integrate the sky spectral template with the given exposure time and add the Poisson noise to the 3D data cube as drawn from the cumulative sky emissions. We also add a Gaussian read noise. Finally, we convolve each 2D slice with a PSF across the $\lambda$-direction. The resulting 3D simulated data cube then enters the observational mask with given slit width and slit angle to derive the output slit spectrum. 

\begin{figure}
  \centering
  \includegraphics[width=\linewidth]{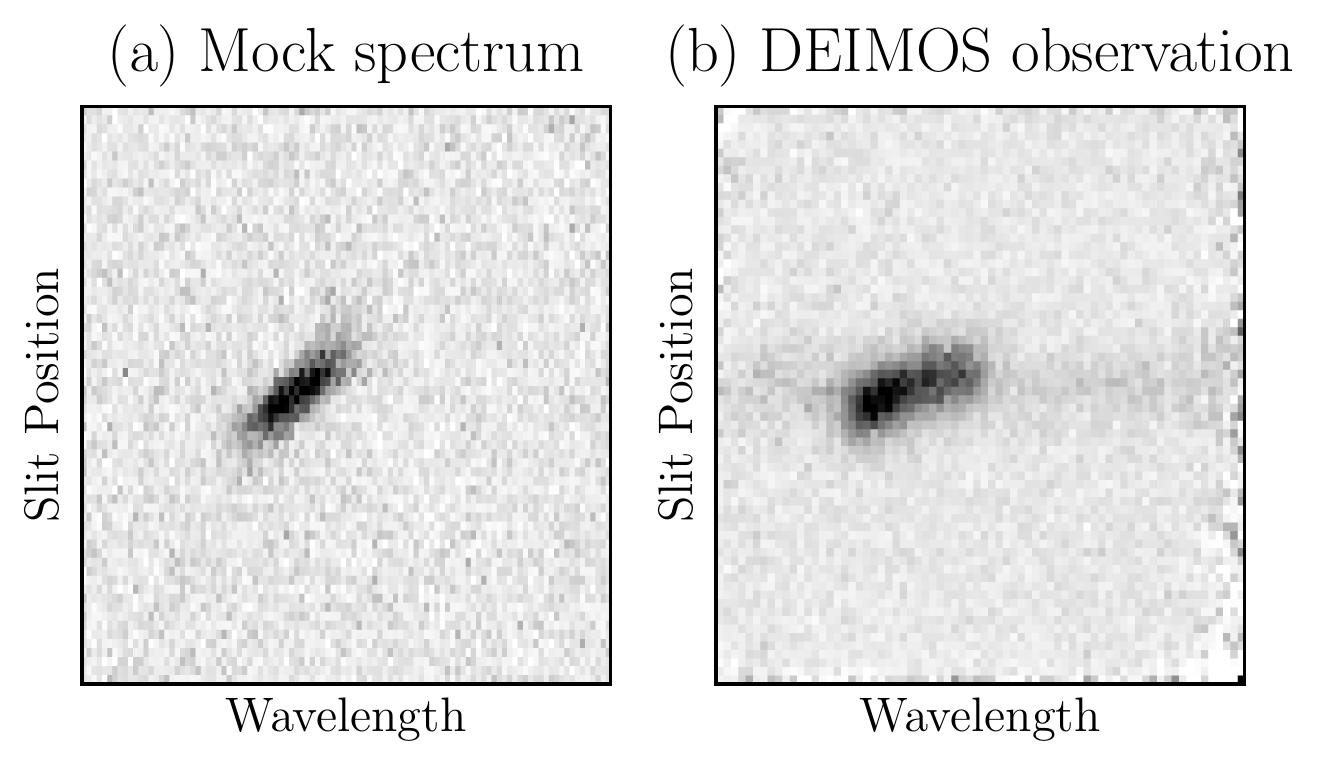}
  \caption{Comparison of mock and observed H$\beta$ emission line. Panel (a) is the spectrum from our mock data generator, and panel (b) is the observed spectrum from the DEIMOS instrument.}
  \label{fig:mock_data_comparison}
\end{figure}

The above procedure can generate a slit spectrum for any number of emission lines in the wavelength range of interest. In Fig. \ref{fig:mock_data_comparison} 
 we show a comparison of a simulated H$\beta$ spectrum and an observation from DEIMOS archival data.

\begin{table*}
    \centering
    \begin{tabular}{cllll}
    \toprule
    Parameter & Description & Fiducial & Prior & Units\\
    \midrule
    % \decimalcolnumbers
    $\gamma_{+}$ & Shear component & 0.05 & $\mathcal{U}$(-0.2, 0.2) & -\\
    $\gamma_{\times}$ & Shear component & 0.05 & $\mathcal{U}$(-0.2, 0.2) & -\\
    $v_{\mathrm{circ}}$ & Maximum circular velocity & 200.0 & $\mathcal{N}(\log \hat{v}_{\mathrm{circ}}, \sigma_{\mathrm{TF}})$ & kms$^{-1}$\\
    $v_0$ & Galaxy systemic velocity & 10.0 & $\mathcal{U}$(-100, 100) & kms$^{-1}$\\
    $\sin i$ & Galaxy inclination & Varied & $\mathcal{U}$(0, 1) & -\\
    $r_{\mathrm{hl}}^{\mathrm{image}}$ & Image half-light radius & 0.75 & $\mathcal{U}$(0.15, 2) & arcsec\\
    $r_{\mathrm{hl}}^{\mathrm{spec}}$ & Spectrum half-light radius & 0.75 & $\mathcal{U}$(0.15, 2) & arcsec\\
    $r_{\mathrm{vscale}}$ & Velocity scale radius & 0.5 & $\mathcal{U}$(0.1, 2) & arcsec\\
    $\theta_{\mathrm{int}}$ &Intrinsic galaxy position angle & $\pi/3$ & $\mathcal{U}$(0, $2\pi$) & radians\\
    $I_0$ & Central brightness & - & $\mathcal{U}(0, 10^3)$ & arbitrary units\\
    $bkg$ & Background & - & $\mathcal{U}(-10, 10)$ & arbitrary units\\
    $q_z$ & Edge-on aspect ratio & 0.2 & Fixed & - \\
    $\sigma_\lambda^{\mathrm{int}}$ & Intrinsic emission line width & 0.01 & Fixed & nm\\
    \bottomrule
\end{tabular}
\label{table:pars}
\caption{List of model parameters along with the fiducial values and priors used. All parameters except for $I_0$ and $bkg$ are shared among the mock generator and the fast forward model.}
\label{table:fit_pars}

\end{table*}

\subsection{Spectrum Fast Forward Model}
\label{sec:forward_model}
The mock spectrum generator is computationally too expensive to be used as a model in KL parameter inference, which requires thousands of model evaluations for each source galaxy. Thus we also develop a fast forward model that skips convolutions and only evaluates points that contribute to the slit measurement.

We model the galaxy as a 2D grid and apply a slit mask that can be placed at an arbitrary angle w.r.t. the galaxy's major axis. This limits our computations to points within the slit. We map these points from image to face-on plane by inverting Eq. (\ref{eq:coord_mapping}) and compute the LoS velocity using the arctan rotation curve given by Eq. (\ref{eq:vel_profile}).

Similar to the mock spectrum generator, we assume a Gaussian profile for the emission line and calculate the peak wavelength for each position based on the LoS velocity. The width of the emission line is a combination of the intrinsic spread and the instrument resolution (Eq. (\ref{eq:sigma2})). To each pixel within the slit, we assign the emission line intensity from a PSF-convolved galaxy image with half-light radius $r_{\mathrm{hl}}^{\mathrm{spec}}$. The galaxy image used for prescribing the emission line intensity has a different half-light radius since the image and spectrum are sourced by different emission processes (stellar continuum and line emission, respectively); hence having a separate scale radius for each is physically more realistic. 

After assigning line intensity to each pixel, the brightness profile is multiplied by a nuisance parameter $I_0$ that determines the central brightness. Combining the brightness profile with the line emission results in a 3D model cube with two spatial axes (corresponding to the slit dimensions) and one wavelength axis. Finally, by integrating this data cube across the slit width, we obtain the 2D spectrum. The grid resolution for the integral is chosen based on the instrument for which the spectrum is observed/generated. We also add a flat background noise $bkg$ to account for any residual noise after sky subtraction.

\section{Simulated KL Inference}
\label{sec:kl_measurement}
In this section, we simulate KL parameter inference by analyzing mock data with the fast forward model.
To differentiate between model biases and noise biases, we first consider noiseless mock data. After obtaining unbiased shear estimates in this idealized setup, we characterize shear bias and shape noise for an ensemble of noise realizations.

\subsection{Mock Data Characteristics}
Using the mock generator, we simulate data consisting of a galaxy image and two 2D spectra from orthogonal slits for each galaxy. The instrument characteristics used to simulate mock data, shown in Table \ref{tab:DEIMOS_pars}, are based on Keck DEIMOS, a multi-slit imaging spectrograph used to carry out the DEEP2 Galaxy Redshift Survey. DEEP2 has publicly available data with high spectral resolution and accurate target redshifts \citep{Newman_2013}, which makes it a promising candidate for a KL measurement with real observations.

We simulate images with a signal-to-noise ratio (SNR) of 100. For the spectrum, we simulate the H$\alpha$ emission line at $z=0.4$ with a SNR of 30, representative of archival DEIMOS data. We define the emission line SNR using the corresponding 1D spectrum $f_\lambda$ and wavelength-dependent sky variance $\sigma_\lambda^2$
\begin{equation}
    \text{Emission line SNR} = \sqrt{\sum_\lambda \frac{f_\lambda^2}{f_\lambda + \sigma_\lambda^2}}.
\end{equation}

\subsection{Fitting Methodology}
\label{sec:fit_method}
The simulated mock data vector is fitted using the image and spectrum generated from the fast forward model.
We use the Markov Chain Monte Carlo (MCMC) package
\texttt{Emcee} \citep{emcee} to explore the multi-dimensional parameter space and find best fit values based on the likelihood $\mathcal{L}$, given by
\begin{align}
\label{eq:log_like}
    \ln\mathcal{L}(\boldsymbol \Theta) &= -\frac{1}{2}\Big[\sum_{j=1}^{2} \sum_{k} \frac{\left[D_{\mathrm{spec},j}-M_{\mathrm{spec},j}(\boldsymbol \Theta)\right]_k^2}{\sigma_{\mathrm{spec}}^2}  \notag\\
    &+\sum_{k}\frac{\left[D_{\mathrm{im}}-M_{\mathrm{im}}(\boldsymbol \Theta)\right]_k^2}{\sigma_{\mathrm{im}}^2}
    + \Big(\frac{\log v_{\mathrm{circ}}(\boldsymbol \Theta)-\log \hat{v}_{\mathrm{circ}}}{\sigma_{\mathrm{TF}}} \Big)^2 \Big].
\end{align}
Here, $D_{\mathrm{im}}$ is the mock image and $D_{\mathrm{spec}, j}$ is the mock spectrum with the index $j$ corresponding to the two slit angles. Similarly, $M_{\mathrm{im}}$ and $M_{\mathrm{spec}, j}$ are the model image and spectra at parameter values $\boldsymbol\Theta$. The summation $k$ is over all the pixels for both spectrum and image. The image noise is quantified by a scalar $\sigma_{\mathrm{image}}$ and the 2D spectrum variance $\sigma^2_{\mathrm{spec}}$ uses the sky template described in section \ref{section:tfCube}. The last term in the likelihood is the TFR-based log-normal prior for the maximum circular velocity centered at $\log \hat{v}_{\mathrm{circ}}$ and with scatter $\sigma_{\mathrm{TF}}$. For KL inference on a per-galaxy basis, as in this paper, the TFR parameters are held fixed based on observed constraints. However, we note that these could become hyperparameters when fitting a large number of galaxies simultaneously.

In Table \ref{table:fit_pars} we list the parameters of our model. For each parameter, we report the fiducial value used in mock data, the prior if the parameter is varied in the MCMC routine, and the appropriate units. As described in section \ref{sec:kl_theory}, we use the TFR to constrain the maximum circular velocity. The scatter for the TFR prior is set to $\sigma_{\mathrm{TF}}=0.08$ dex (see Table 4 in \citealt{Miller_2011}). We limit the photometric and spectroscopic galaxy half-light radii to the range (0.4, 1.4), based on R-band galaxy sizes from the DEEP2 photometric catalog \citep{Coil_2004}; we do not expect the prior range for different bands to vary significantly. We use uniform priors for all other parameters, with bounds set to exclude unphysical values.

\begin{table}
\centering
\begin{tabular}{ll}
\toprule
Description &Value\\
\midrule% \decimalcolnumbers
Gain & 1.0 e$^-$/ADU\\
PSF FWHM & 0.5 arcsec\\
Spectral Resolution & 5000 \\
Read noise & 3 e$^-$/pix\\
Slit width & 1.0 arcsec\\
Spatial pixel scale & 0.1185 arcsec/pix \\
Spectral pixel scale & 0.033 nm/pix\\
Throughput & 0.29\\
\bottomrule
\end{tabular}
\caption{Instrument characteristics used in mock generator.}
\label{tab:DEIMOS_pars}
\end{table}

\begin{figure}
  \centering
  \includegraphics[width=\linewidth]{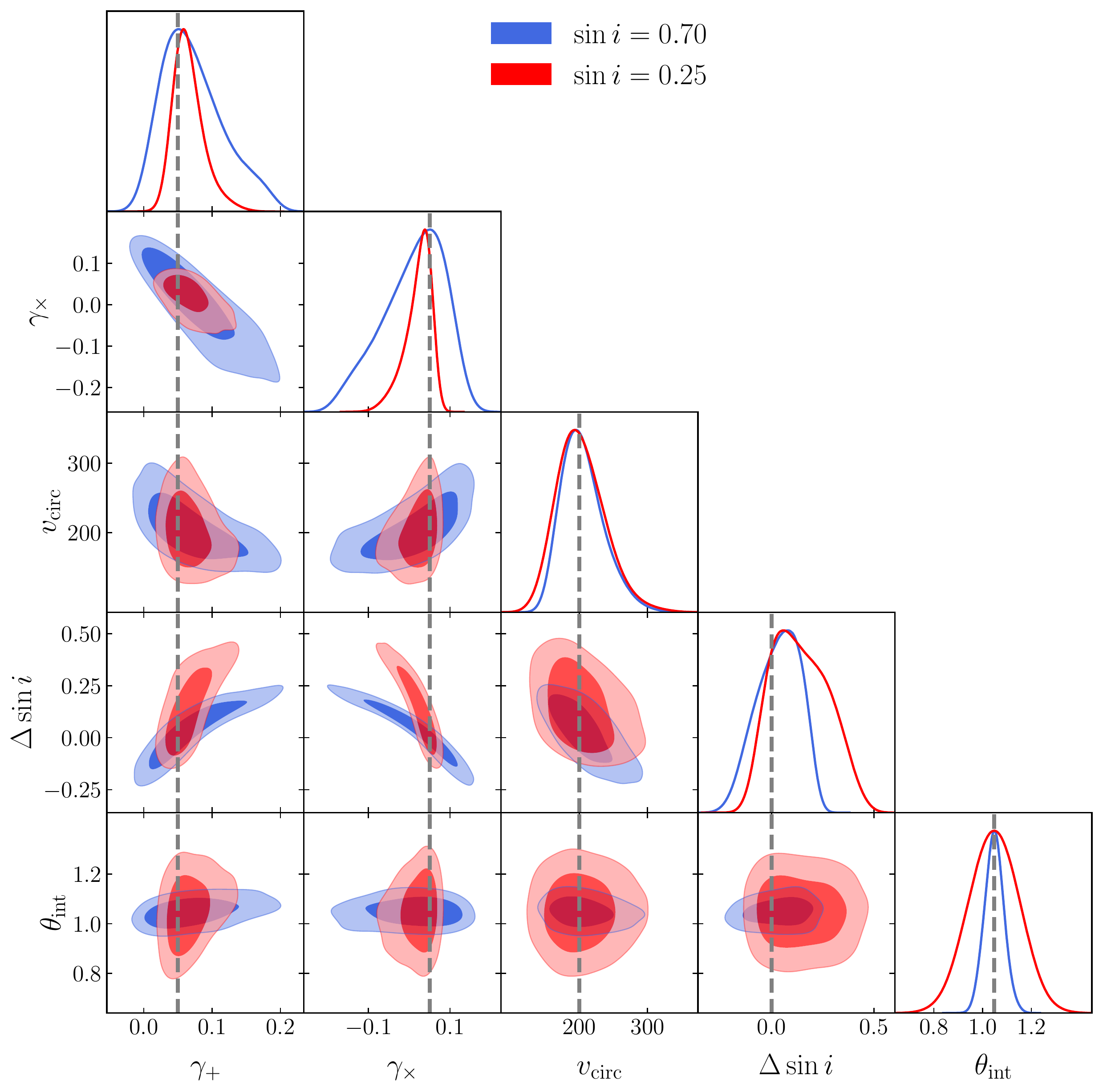}
  \caption{Posterior distribution for shear and model parameters for different galaxy inclinations. The constraints are derived from noiseless mock data with image and emission line SNR set to 100 and 30 respectively. See Fig. \ref{fig:triangle_plot_full} for constraints on all fit parameters.}
  \label{fig:triangle_plot}
\end{figure}

In total we fit for 11 parameters by running 100 walkers for 30,000 steps and discard 70\% of the samples as burn-in.

\subsection{Validation on Noiseless Mocks}
\label{sec:validation}
\begin{figure*}
    \centering
    \includegraphics[width=\linewidth]{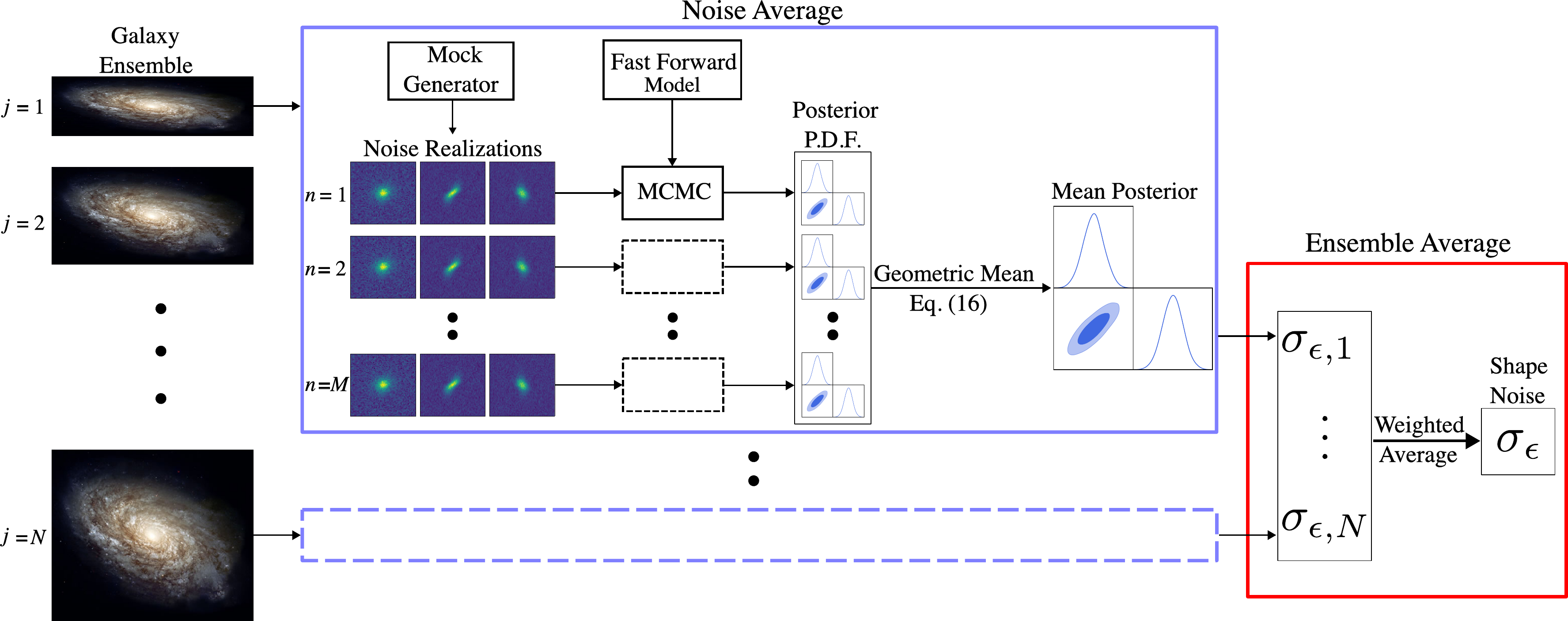}
    \caption{Schematic of the ensemble-averaged shape
noise estimation procedure. The galaxy ensemble consists of $N=20$ galaxies with inclination values sampled uniformly in $\cos i$. The blue box outlines the noise average procedure: For each inclination we generate several noise realizations using the mock generator which are then individually fitted by the fast forward model; the marginalized posteriors from the MCMC fits are combined in mean parameter posteriors. The red box indicates the ensemble average, which combines the shear posteriors of individual galaxies into the galaxy-averaged shape noise.}
    \label{fig:pipeline_overview}
\end{figure*}

As a first test of the KL inference pipeline, we fit the 11-parameter fast forward model to noiseless mock data at the baseline SNR of 30 for the emission line and SNR of 100 for the image. Figure \ref{fig:triangle_plot} shows posteriors for two galaxy inclinations from MCMC fits of this noiseless mock data. We recover the fiducial values of all fit parameters with $<1\sigma$ bias. More importantly, combining imaging and spatially resolved spectroscopy allows us to obtain unbiased shear estimates with a mean shear uncertainty of $\approx 0.03$ ($\approx 0.06$) for $ \sin i =0.25$ ($\sin i =0.7$). This example also illustrates the strong dependence of shear uncertainty on galaxy inclination, with the uncertainty in the two cases differing by a factor of $\approx2$. This is because the shape distortion (Eq. (\ref{eq:e_obs})), the misalignment of the velocity field (Eq. (\ref{eq:theta_obs})) and the offset of the LoS velocity from TFR (Eq. (\ref{eq:vmajor})) are larger for a galaxy that is more face-on i.e., inclined at a smaller angle w.r.t. the LoS. Thus lower inclination galaxies contain more photometric and kinematic information, providing tighter constraints.

\subsection{Results}
To characterize the expected performance of the KL shear inference pipeline, we next quantify its bias and variance in the presence of additive (photon) noise in the data. As the parameter inference is non-linear in pixel values, pixel noise may lead to biases in the inferred model parameters. For each galaxy $j$ (specified by galaxy properties, observational characteristics, and input shear), we generate $M$ noise realizations to verify that shear estimates are unbiased after averaging over noise realizations. For the results presented in this paper we choose $M=20$. For each noise realization $n$ we perform KL inference to obtain the parameter posterior $p_{j,n}(\boldsymbol \Theta)$. The different noise realizations are combined by taking the geometric mean of the marginalized posteriors
\begin{equation}
\label{eq:geometric_mean}
    \bar{p}_j(\Theta_\alpha) = \left(\prod_{n=1}^M p_{j,n}(\Theta_\alpha)\right)^{1/M},
\end{equation}
where $\bar{p}_{j}$ is the noise realization-averaged marginalized posterior for parameter $\Theta_\alpha$. 

Then the mean $\mu_j$ and the standard deviation $\sigma_j$ of parameter $\Theta_\alpha$ are computed from the mean posteriors as
\begin{align}
    \mu_j(\Theta_\alpha) &= \int \bar{p}_{j}(\Theta_\alpha) \Theta_\alpha d\Theta_\alpha, \\
    \label{eq:est_std_dev}
    \sigma_j(\Theta_\alpha) &= \sqrt {\int \bar{p}_{j}(\Theta_\alpha) (\Theta_\alpha-\mu_j(\Theta_\alpha))^2 d\Theta_\alpha}.
\end{align}
This procedure for obtaining noise-averaged parameter posteriors for each galaxy/set of input parameters is summarized by the blue box in Fig.~\ref{fig:pipeline_overview}.

\begin{figure}
  \centering
  \includegraphics[width=\linewidth]{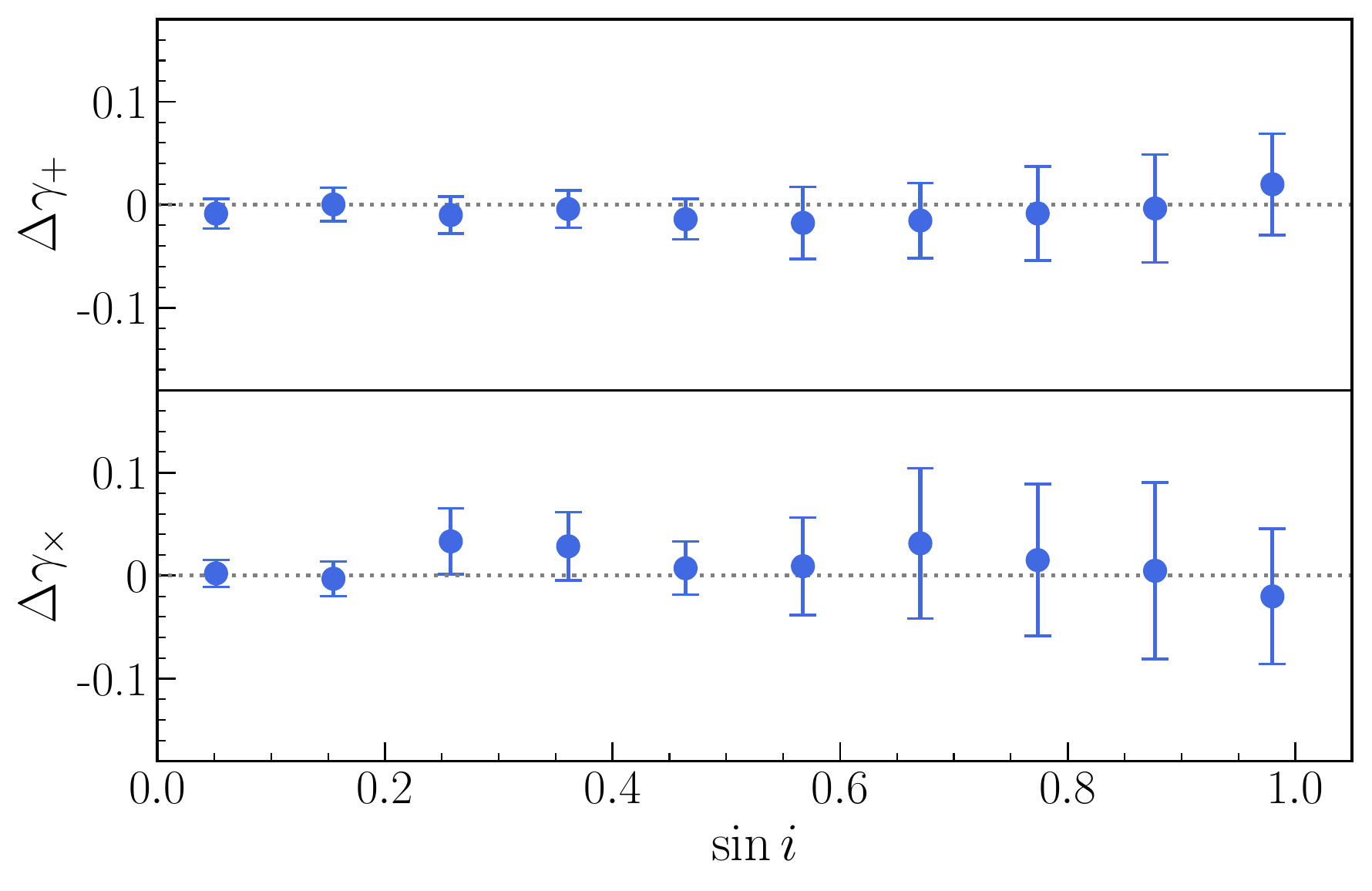}
  \caption{Bias in shear components as a function of galaxy inclination. Averaged over galaxy orientation, the mean bias per shear components is about $\langle \Delta\gamma_{+, \times} \rangle \approx0.005$.}
\label{fig:shear_bias}
\end{figure}

Using the above procedure, we analyze mock galaxies at different inclination values and estimate shear bias $\Delta \gamma_{+,\times} = \gamma_{+,\times}^{\mathrm{measured}}- \gamma_{+,\times}^{\mathrm{input}}$ as a function of galaxy inclination, shown in Fig. \ref{fig:shear_bias}. Averaging over galaxy orientations, we find an average additive shear bias $\langle \Delta\gamma_{+, \times} \rangle \approx 0.005$, which is small compared to the average shear in small-field KL applications like cluster lensing. For cosmic shear measurements, where typical additive shear requirements are $\Delta \gamma =\mathcal{O}(10^{-4})$ \citep{MasseySystematics}, a large volume of simulations (varying multiple galaxy properties and encompassing a much larger number of noise realizations) will be required.

\begin{figure}
  \centering
  \includegraphics[width=\linewidth]{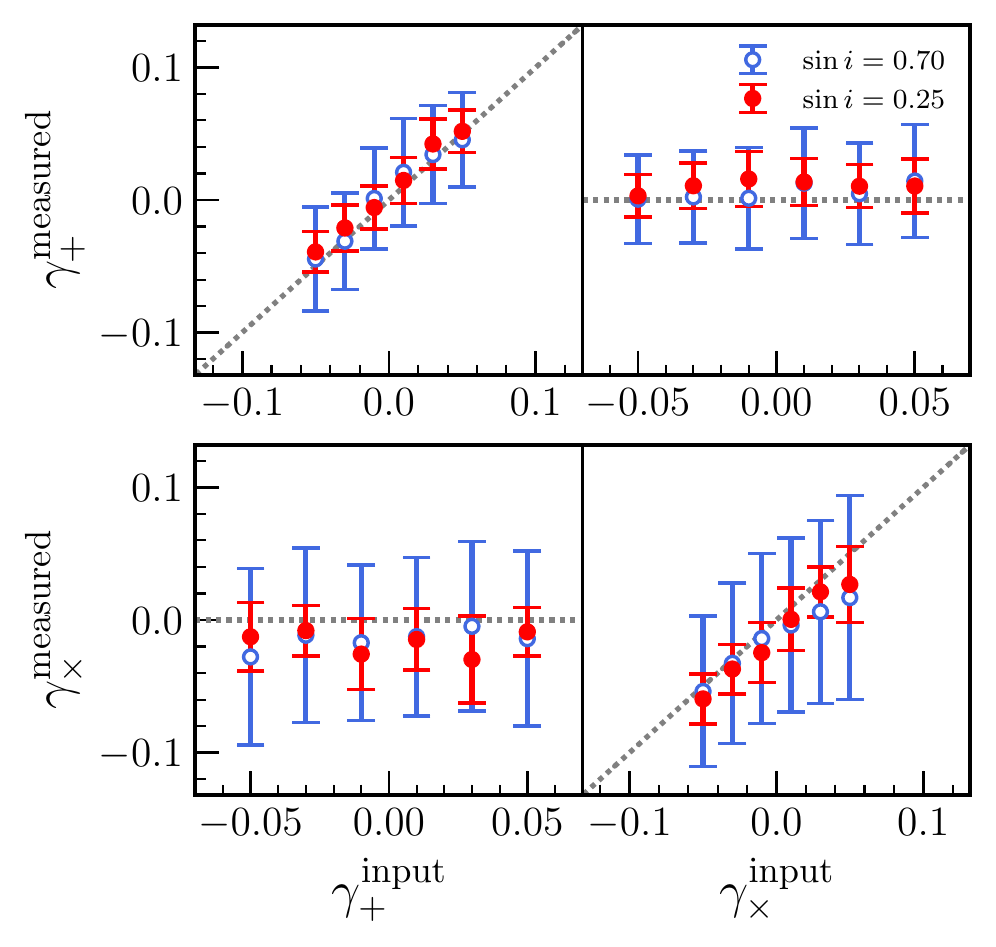}
  \caption{Measured vs. input shear for two galaxy inclinations. We vary the input value of one shear component at a time and fix the other to zero. The left and right columns show the measured shear when $\gp$ and $\gc$ are varied, respectively. The gray dotted line represents the input shear value. Our KL pipeline infers both shear components without significant contamination between shear components or multiplicative bias.}
\label{fig:input_shear_bias}
\end{figure}
While a detailed characterization of multiplicative and additive shear biases is beyond the scope of this paper, we next test the performance of KL inference as a function of input shear. Specifically, we generate a family of galaxies at the baseline input parameters with $\mathrm{SNR}=30$, varying the input shear one component at a time: $\left(\gamma_+ =\{-0.05,-0.03,-0.01,0.01,0.03,0.05\},\gamma_\times =0\right)$ and $\left(\gamma_+ =0,\gamma_\times =\{-0.05,-0.03,-0.01,0.01,0.03,0.05\}\right)$. The results of this test are shown in Fig.~\ref{fig:input_shear_bias} for two different galaxy inclinations. Defining multiplicative shear biases as
\begin{gather}
    \begin{aligned}
    \gamma_{+, \times}^{\mathrm{measured}} &= (1+m_1)\gamma_{+, \times}^{\mathrm{input}}, &
    \gamma_{+, \times}^{\mathrm{measured}} &= m_2\gamma_{\times, +}^{\mathrm{input}},
\end{aligned}
\end{gather}
we find $\langle m_1 \rangle = -0.05 \pm 0.06$ and $\langle m_2 \rangle=0.02 \pm 0.08$. KL inference appears to separate the two shear components well, with no significant contamination of the shear component held zero ($m_2$, upper right and lower left panels). While this test with limited  noise realizations and galaxy inclinations provides insufficient statistics to quantify multiplicative or additive shear biases, the agreement between input shear and measured values is encouraging for KL cosmic shear measurements.

As the shape noise is insensitive to effects at the current level of additive bias constraints, we next estimate the KL shape noise for a population of randomly oriented disc galaxies, which are uniformly distributed in $\cos i$.
We generate an ensemble of $N =20$ galaxies uniformly selected in $\cos i$.\footnote{For numerical stability we exclude perfectly edge-on discs and therefore the inclination range for our ensemble is $\cos i \in (0.1, 1)$. As the tests in Fig.~\ref{fig:shear_bias} found consistent results for positive and negative inclination values, we only simulate positive inclination values.}.

For each galaxy $j$, we again generate $M=20$ noise realizations, compute the noise-averaged variance of the shear components ($\sigma_{{\gamma_{+/\times}}, j}$, Eq.~(\ref{eq:est_std_dev})) and the shape noise estimate
\begin{align}
    \sigma_{\epsilon, j} = \frac{\sqrt{\sigma_{{\gamma_+}, j}^2 + \sigma_{{\gamma_\times}, j}^2}}{\sqrt{2}}.
\end{align}
For $M$ noise realizations, the standard error on the shape noise estimate for galaxy $j$ is given by
\begin{align}
    s_{\sigma_{\epsilon, j}} = \frac{\sigma_{\epsilon, j}}{\sqrt{2(M-1)}}.
\end{align}
The ensemble/orientation-averaged shape noise is then calculated with an inverse variance weighting (with weight $w_j =1/ s^2_{\sigma_{\epsilon, j}}$), which yields $\sigma_\epsilon=0.038$.

As described in the discussion of Fig.~\ref{fig:triangle_plot} and Fig.~\ref{fig:shear_bias}, the measurement uncertainty increases as we move toward edge-on systems, which has implications for the KL source sample optimization strategy. In the limit that there are more target disc galaxies in the survey area than one can follow up spectroscopically, targeting lower inclination galaxies will increase the precision of shear measurements, which we explore in Sec. \ref{sec:dependence_sini}.

\section{Shape noise estimates and dependencies}
\label{sec:discussion}
To guide the design of future KL surveys, in this section we characterize the dependence of KL shape noise on different observational characteristics and modeling assumptions.

\subsection{Emission Line SNR}
\begin{figure}
  \centering
  \includegraphics[width=\linewidth]{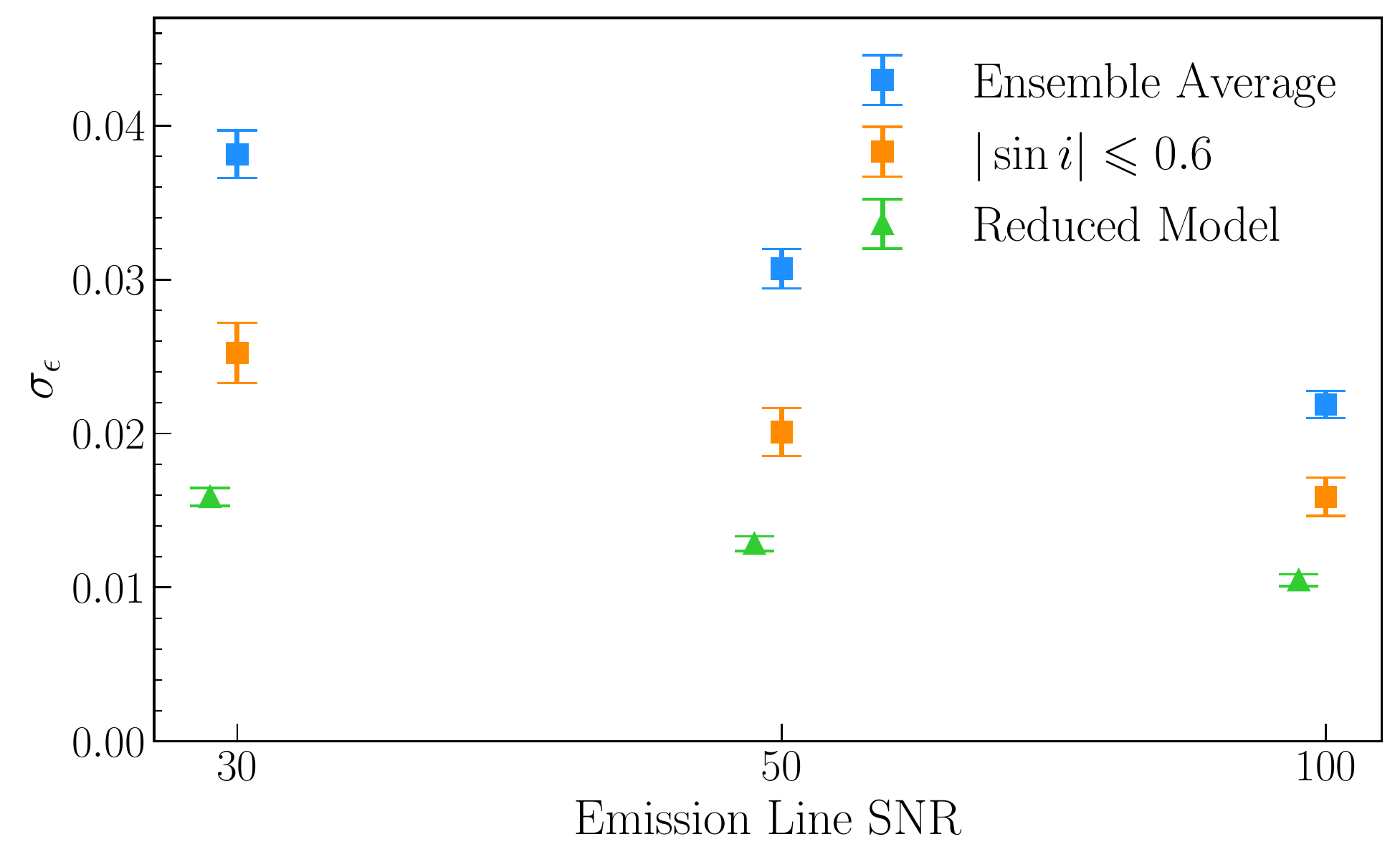}
  \caption{Dependence of estimated shape noise on emission line SNR. The image SNR is set to 100 for all cases. The blue points show the ensemble-averaged shape noise and the orange points the estimate excluding galaxies with inclination $|\sin i|>0.6$. The green triangles indicate the shape noise when we use the fast forward model to both generate and fit simplified mock data with just 6 parameters. The error bars are based on the number of noise realizations used for each galaxy inclination.  Data points are slightly offset along the x-axis for improved readability.}
  \label{fig:shapenoise_vs_snr}
 \end{figure}
 
The signal-to-noise ratio of follow-up spectroscopy is one of the easiest to control but potentially expensive characteristics of KL observations. We characterize the impact of emission line SNR on shape noise in Fig. \ref{fig:shapenoise_vs_snr}. In the $\mathrm{SNR}=30$ case, we estimated the ensemble-averaged shape noise to be $\sigma_\epsilon=0.038$. We find that increasing the $\mathrm{SNR}$ of spectroscopic observations to $100$ would further reduce the shape noise for KL shear measurements, corresponding to $\sigma_\epsilon\approx 0.022$. As a linear reduction in shape noise (per galaxy) compensates for a quadratic reduction in the number of source galaxies, this finding may motivate optimizing a KL survey towards higher SNR observations of fewer source galaxies. We also show the shape noise estimate for a galaxy sample selection restricted to $|\sin i| \leq 0.6$. 

As the shear of lower inclination systems can be measured with higher precision, the resulting shape noise is significantly smaller than that of the complete ensemble average of all inclination values. In the limit that there are more target disc galaxies in the survey area than one can follow up spectroscopically, restricting the KL source galaxy sample selection to less inclined disc galaxies may optimize the ensemble shape noise.

\subsection{Galaxy Inclination}
\label{sec:dependence_sini}
\begin{figure}
  \centering
  \includegraphics[width=\linewidth]{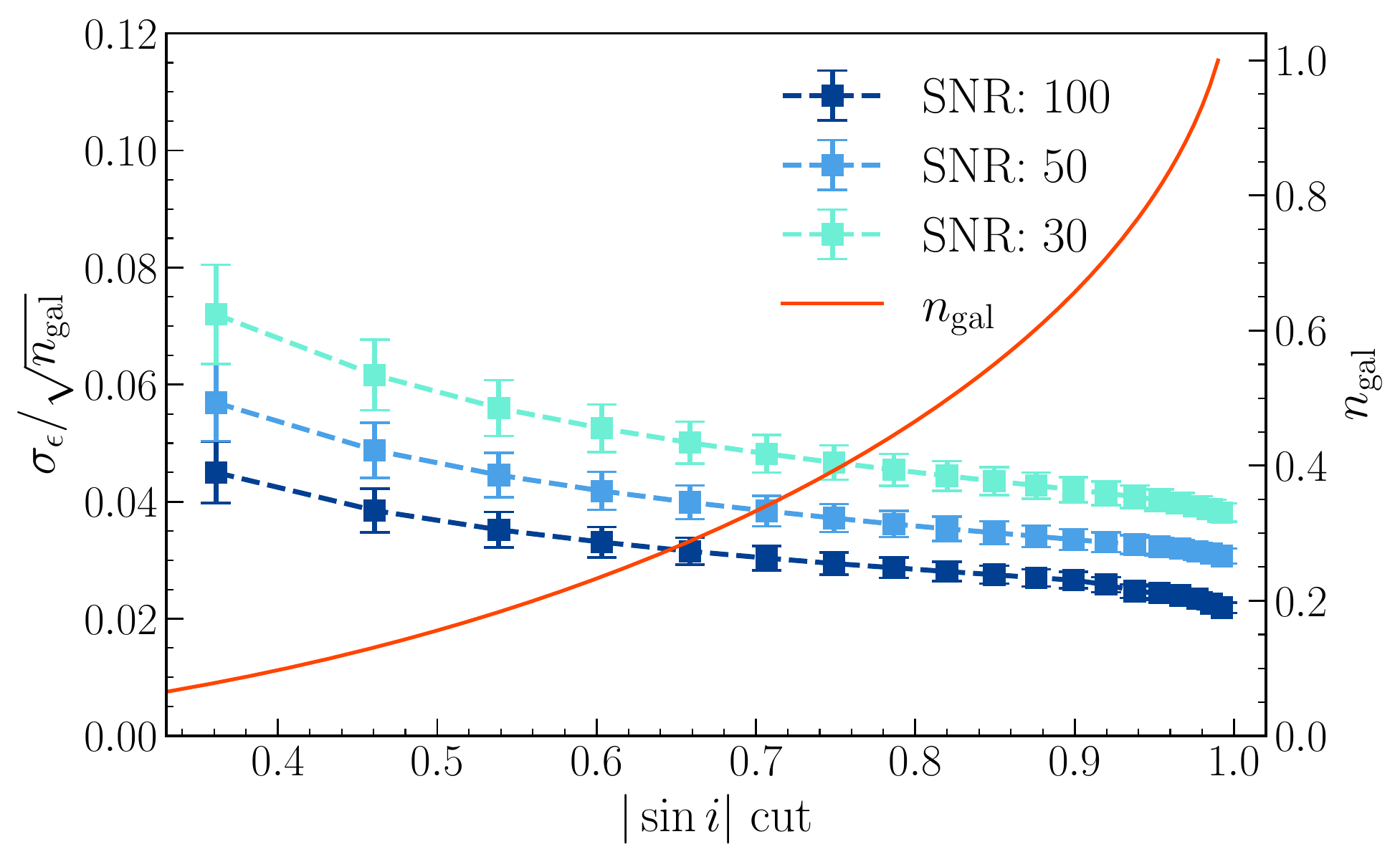}
  \caption{Shape noise estimates for galaxies within a $\sin i$ threshold, weighted by the fraction of randomly oriented galaxies within that inclination threshold, $n_{\mathrm{gal}}$ (red line, shown on the right y-axis). While highly inclined galaxies provide weaker shear constraints per galaxy, relaxing the inclination threshold reduces the average shape noise by increasing the number of source galaxies.}
  \label{fig:shapenoise_vs_sinicut}
\end{figure}
While an inclination threshold is advantageous in the limit that there are more target disc galaxies in the survey area than one can follow up spectroscopically, such a restriction reduces the source number density when the target disc galaxy sample is limited. Accounting for both of these effects, Fig. \ref{fig:shapenoise_vs_sinicut} shows the shape noise estimates for galaxy samples within an inclination threshold, weighted by the fraction of randomly oriented source galaxies within that inclination threshold, $n_{\mathrm{gal}}$.

While increasing the sample size with higher inclination galaxies improves the constraining power, these gains are slow beyond $|\sin i| \gtrsim 0.6$, which corresponds to just $n_{\mathrm{gal}}\approx 20\%$ of the available target sample. For a fixed number of spectra taken, targeting less-inclined galaxies over a wider survey area would improve the KL survey strategy.
 
\subsection{Scatter of the Tully-Fisher Relation}
\begin{figure}
  \centering
  \includegraphics[width=\linewidth]{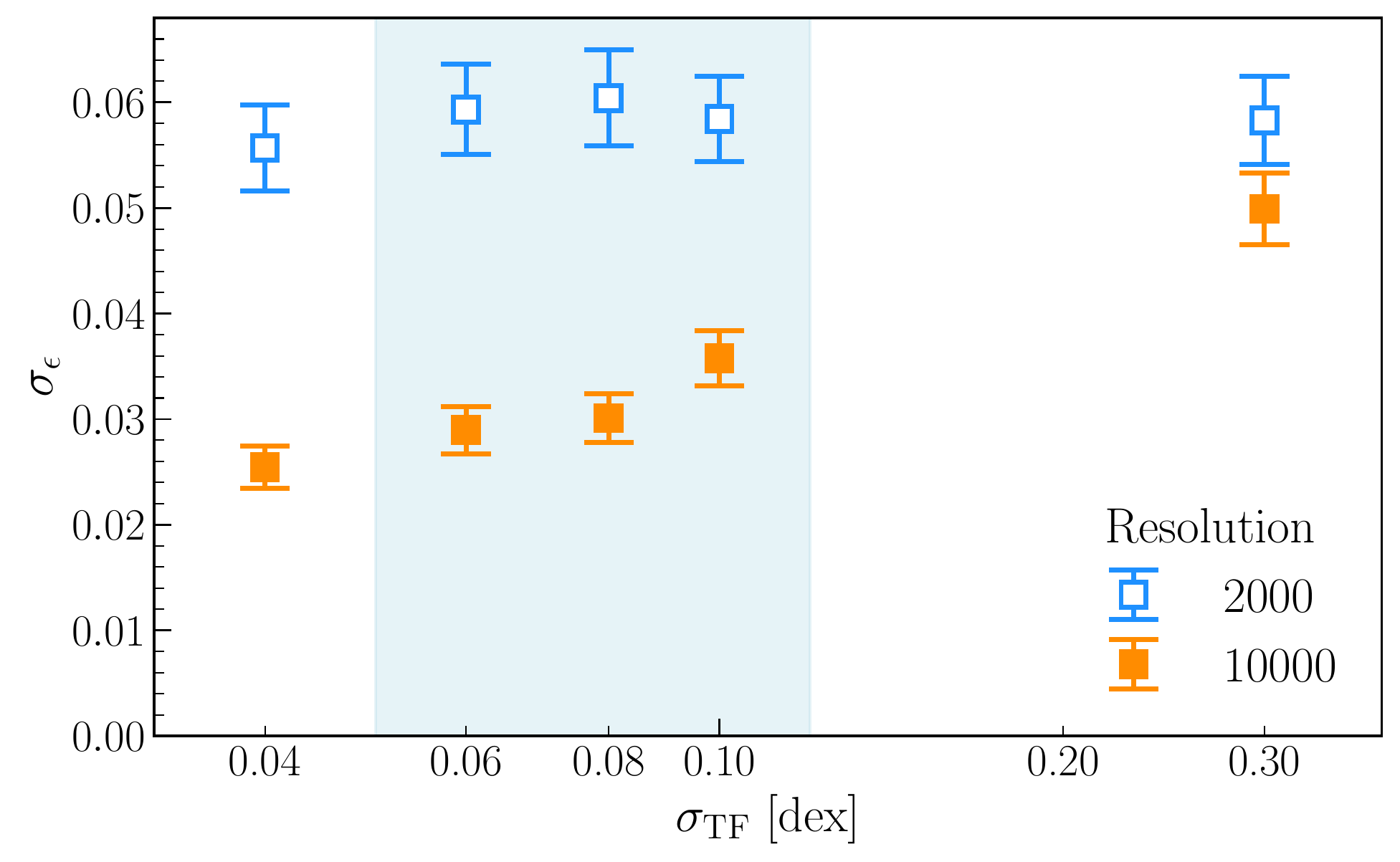}
  \caption{Shape noise as a function of scatter in the TFR. The blue and orange points show estimates for two different spectral resolutions, $R=2000$ and $R=10000$ respectively.  
  The shear uncertainty appears mostly insensitive to the TFR scatter. The blue shaded region shows the observed range of $\sigma_{\mathrm{TF}}$. The shape noise appears mostly insensitive to the TFR scatter.}
  \label{fig:sigma_gplus_vs_sigmaTF}
\end{figure}

In the simple shear estimator picture described in Sect.~\ref{sec:kl_theory}, the shape noise depends directly on the uncertainty of the circular velocity estimate, which is a combination of the observational velocity measurement uncertainty (inversely proportional to spectral resolution and SRN) and the TFR scatter. At spectral resolutions where the uncertainty in the circular velocity estimate is limited by the TFR scatter, one expects the shear uncertainty to be proportional to the TFR prior. This trend is reflected in Fig. \ref{fig:sigma_gplus_vs_sigmaTF}, where the shape noise at $R=10000$ appears to be proportional to $\sigma_{\mathrm{TF}}$. However, we find the shape noise for $R=2000$, $\mathrm{SNR}=30$ to be largely insensitive to the TFR scatter. This indicates that the underlying parameter degeneracies for KL inference in the 11-dimensional parameter space are not easily traceable and corresponding sensitivity studies need to be carried out for each KL survey design, taking into account detailed observational characteristics beyond spectral resolution.

\subsection{Impact of Modeling Assumptions}
In a separate experiment, we use a simplified fast forward model first to generate the mock data and then as model in the fit, reducing the model complexity to just six key parameters $(\gp, \gc, v_{\mathrm{circ}}, \sin i, \theta_{\mathrm{int}}, I_0)$. This reduced model eliminates differences between the mock data and the model, also known as model misspecification, and decreases the parameter volume. As shown by the green triangles in Fig.~\ref{fig:shapenoise_vs_snr}, these two factors result in a significantly smaller shape noise at the low signal-to-noise end. It is important to note that the estimated shape noise is sensitive to modeling assumptions and complexity and that future work should optimize the model space and priors.

Throughout the analyses presented here, we have assumed galaxies to be perfectly round discs. In practice, galaxies have not only disc and bulge components but also knots of star formation that add complexity to the galaxy morphology and kinematic structure. Other astrophysical systematics include variations in the disc scale height, which can be degenerate with shear, and dust driven biases in the TFR.
Hence, KL inference with real galaxies may require additional model complexity, which will increase the uncertainty of shear measurements. We will characterize the impact of these astrophysical systematics in future work.

\section{Conclusions}
\label{sec:conclusion}
In this paper we develop and test a KL shear inference pipeline and use this pipeline to map out realistic shape noise levels for the KL technique. We build a realistic mock data set for galaxy images and spectrum closely resembling the combination of high-resolution imaging and Keck DEIMOS spectroscopy. Our mock data account for numerous observational and instrument effects e.g.,  atmospheric transmission, sky emissions, PSF, and detector read noise to ensure sufficient realism. We further develop a fast forward model that infers the KL signal from our realistic mock data. Compared to the mock data, the forward model makes several simplifications, enabling faster likelihood evaluations suitable for parameter inference.

We show that our KL inference pipeline can robustly recover the input shear even when fitting 11 model parameters for noiseless and noisy mock data. Averaging over a population of inclined discs, we estimate the KL shape noise to be $\sigma_\epsilon=0.038/0.022$ at $\mathrm{SNR}=30/100$ for a single emission line. This already represents an order of magnitude improvement over the noise level of traditional WL, and in future analyses we expect further improvements from fitting multiple lines simultaneously.

We find that less-inclined galaxies provide significantly stronger shear constraints per galaxy; the ensemble shape noise $\sigma_\epsilon / \sqrt{n_\mathrm{gal}}$
saturates at $\sin i \approx 0.6$ even though a randomly oriented galaxy population is distributed towards the high-inclination end. We note that higher quality spectroscopic data can further reduce the KL shape noise. Thus, an optimized KL survey strategy should prioritize quality spectra of low-inclination galaxies instead of maximizing the overall number density.   

Within the observed range of the TFR scatter, we find that the assumed value does
not significantly impact the shape noise. However this trend may change depending on instrument properties, data characteristics or in the presence of astrophysical systematics.

The KL inference pipeline presented here lays the groundwork for a series of future papers in which we will characterize the impact of astrophysical systematics (complex morphologies and kinematic substructure; variations in the TFR) and perform a pilot measurement on real data.

More broadly, KL is unaffected by some of the key systematics affecting WL. Spectroscopic observations ensure that source redshift uncertainties are insignificant, and possible IA contaminations of disc galaxies are suppressed because KL infers the unlensed galaxy shape. Photometric shape measurement uncertainties are less of a concern given the high SNR of the imaging data. 

Our findings support that KL will enable competitive constraints with an order of magnitude smaller source sample, which makes a KL observational program promising with existing spectroscopic instruments already. Given the field-of-view limitations of suitable spectrographs, cluster lensing and galaxy-galaxy lensing measurements are immediate science applications. 

These applications will significantly broaden in the near future with the advent of data from wide-field spectroscopic surveys like the Dark Energy Spectroscopic Instrument survey (DESI\footnote{\href{https://www.desi.lbl.gov/}{\nolinkurl{https://www.desi.lbl.gov/}},} \citealt{DESI_2013}), DESI II \citep{DESI-II}, the Subaru Prime Focus Spectrograph survey (PFS\footnote{\href{https://pfs.ipmu.jp/index.html}{\nolinkurl{https://pfs.ipmu.jp/index.html}}}, \citealt{Takada_PFS}), the 4-metre Multi-Object Spectroscopic Telescope survey (4MOST\footnote{\href{https://www.4most.eu/cms/}{\nolinkurl{https://www.4most.eu/cms/}}}, \citealt{4MOST}), \textit{Roman} and \textit{Euclid}. While KL cosmic shear is the most direct application for cosmology, the feasibility of this measurement must be examined in the context of specific instrument capabilities and survey strategies.

\section*{Acknowledgements}
This work was supported by NASA ROSES ADAP 20-ADAP20-0158. EK and PRS were supported in part by the David and Lucile Packard Foundation and an Alfred P. Sloan Research Fellowship.

The analyses in this work were carried out using the High Performance Computing (HPC) resources supported by the University of Arizona TRIF, UITS, and RDI and maintained by the UA Research Technologies Department.

%%%%%%%%%%%%%%%%%%%%%%%%%%%%%%%%%%%%%%%%%%%%%%%%%%
\section*{Data Availability}
The data underlying this work will be shared on reasonable request to the corresponding author.


\bibliographystyle{mnras}
\bibliography{ref.bib} % if your bibtex file is called example.bib

%%%%%%%%%%%%%%%%% APPENDICES %%%%%%%%%%%%%%%%%%%%%

\appendix
\section{Posterior for all model parameters}
\begin{figure*}
  \centering
  \includegraphics[width=\linewidth]{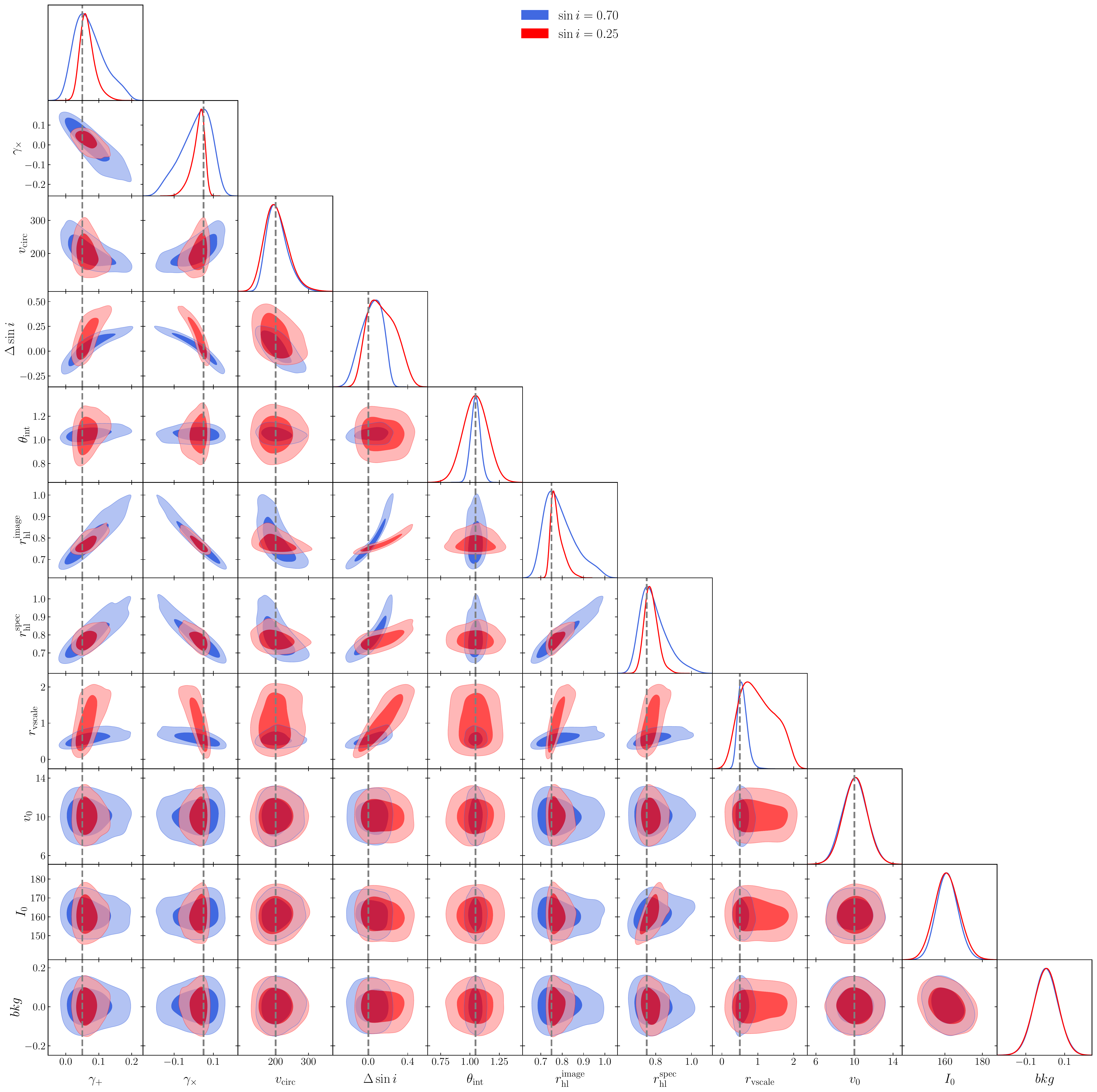}
  \caption{Complete posterior distribution for model parameters for two different galaxy inclinations derived from noiseless mock data.}
  \label{fig:triangle_plot_full}
\end{figure*}

See Fig. \ref{fig:triangle_plot_full} for the full version of Fig. \ref{fig:triangle_plot}.

% Don't change these lines
\bsp	% typesetting comment
\label{lastpage}
\end{document}